\documentclass[twocolumn]{aastex631}
\usepackage{graphicx}
\usepackage{comment}

\newcommand{\Msun}{M$_\odot$}
\newcommand{\teff}{$T_\mathrm{eff}$}

\begin{document}

\title{Astrophysical Properties of the Sirius Binary System Modeled with MESA\footnote{Released 
on March, 1st, 2021}}

\author[0000-0002-0786-7307]{Momin Y. Khan}
\affiliation{Department of Physics, Baylor University, Waco, TX 76798-7316} 
\email{momin\_khan1@baylor.edu}

\author[0000-0001-7010-7637]{Barbara G. Castanheira}
\affiliation{Department of Physics, Baylor University, Waco, TX 76798-7316}

\shorttitle{Modeling Sirius System}
\shortauthors{Khan et al.}

\begin{abstract}

Sirius is the brightest star in the night sky and, despite its proximity, this 
binary system still imposes intriguing questions about its current 
characteristics and past evolution. \cite{2017ApJ...840...70B} published decades 
of astronometric measurements of the Sirius system, determining the dynamical 
masses for Sirius A and 
B, and the orbital period. 
We have used these determinations, combined with photometric determinations for luminosity and spectroscopic determinations of 
effective temperature (\teff) and metallicity, to model the evolution of the 
Sirius system using MESA (Modules for Experiments in Stellar Astrophysics). 
We have constructed a model grid calculated especially for this system and were able to obtain, for Sirius B, a progenitor mass of $6.0 \pm 0.6$\,\Msun ~, yielding a white dwarf mass of $1.015 \pm 0.189$\,\Msun. Our best determination for age of the system is $203.6 \pm 45$\,Myr with a metallicity of 0.0124. We have compared our best fit models with the ones computed using TYCHO, YREC, and PARSEC, establishing external uncertainties. 
Our results are consistent with the observations and 
support a non-interacting past.  
%$6.0 \pm^{0.2}_{0.5}$\,\Msun
\end{abstract}

%% https://astrothesaurus.org
\keywords{Stellar ages(1581) -- Stellar evolution(1599) -- Stellar mass functions(1612) -- Visual 
binary stars(1777) }

\section{Introduction} \label{intro}
The Sirius binary system is one of the closest binaries to Earth at only 
$2.64\pm0.01$\,pc \citep{2007A&A...474..653V}. Early observations of the astrometric perturbations of Sirius A led to the discovery of an unseen companion 
\citep{1844MNRAS...6R.136B}, which would become the prototype of the white dwarf 
stars, Sirius B. \cite{1915PASP...27..236A} first noted the similarities of the 
spectra of Sirius A and B despite the different values for masses and 
luminosities of these stars. Sirius B and subsequent stars with similar 
characteristics were called white dwarfs \citep{1922PASP...34..356L}, and 
studied as a class by \cite{1944AJ.....51...13R}. The vast majority of white 
dwarfs are observed to be single stars, but studies of the Solar neighborhood 
indicated that approximately 8\% are members of ``Sirius-like'' systems 
\citep{2013MNRAS.435.2077H}, with at least one component of spectral type K or 
earlier. 

The initial--to--final mass relation (IFMR) of stars is fundamental in the 
overall understanding of stellar evolution, from the star 
formation history to the stellar populations and the interpretation of stellar 
luminosity distributions. To determine the IFMR, studies of stellar populations 
estimated empirically this relationship, using star clusters [e.g. 
\cite{2018ApJ...866...21C} and references within]. ``Sirius-like'' systems are 
very useful to the study of IFMR, as well as of the white dwarf mass--radius 
relationship. 

Over the past few decades, much effort has been put into modeling the 
evolution of Sirius A and B \citep{2005ApJ...630L..69L}, but fitting the 
observed parameters of the brightest star in the sky and its companion remains a 
challenging exercise. 

\cite{2017ApJ...840...70B} published a historic analysis of decades of 
astrometric data using the Hubble Space Telescope, photographic observations, 
and measurements dating back to the $19^\mathrm{th}$ century to determine the 
dynamical masses and orbital properties of the Sirius system. We have used their 
mass measurements of $2.063\pm0.023$\,\Msun~and $1.018\pm0.011$\,\Msun~for 
Sirius A and B, respectively, and the orbital period of 50.13\,years of Sirius 
B, with an eccentricity of 0.59, and additional determinations of effective 
temperature (\teff) and metallicity to probe its evolution. We computed a model 
grid for the Sirius system, using the Modules for Experiments in Stellar 
Astrophysics (MESA), an open-source 1D stellar evolution code 
\citep{2011ApJS..192....3P, 2013ApJS..208....4P, 2015ApJS..220...15P, 2018ApJS..234...34P, 2019ApJS..243...10P, 2023ApJS..265...15J}.

In this paper, we present the parameter space of the model grid calculated for 
the stellar evolution of the Sirius system and our best model. We then 
compared our results with the ones from previous studies, establishing external 
uncertainties in our determinations. Our goal is not only to provide a better 
understanding of the evolution of Sirius A and B, but also to shed light on modeling the 
evolution of binary systems as a whole. 

\section{Observational Parameters} \label{observ}

In this section, we discuss the observed physical quantities to determine the parameter space of our 
model grid for the Sirius system. We have made our choices based on how well constrained they are, discussing the internal and external 
uncertainties, and model dependency. 

\subsection{Astrometric Measurements} \label{astrom}

When stars are members of binary systems, astrometric measurements provide the 
most precise mass determinations; it is equivalent of putting the stars on a 
balance scale. Because of the intrinsically robust nature of the astrometric 
determinations, we have fixed the values of mass of Sirius A, determined by 
\cite{2017ApJ...840...70B}, as $2.063\pm0.023$\,\Msun. The current mass of Sirius 
B, $1.018\pm0.011$\,\Msun, also determined by \cite{2017ApJ...840...70B} is our target final mass. 

Besides the mass values, \cite{2017ApJ...840...70B} determined the orbital parameters of the system, establishing that Sirius B orbits around the center of mass of the system in a 50.13 year period, with an eccentricity of 0.59. We have input these values in the calculations of our models. 

Based on observations of stellar winds for Sirius A \citep{1995A&A...302..899B}, 
the mass loss rate derived from the MgII lines is between $2\times10^{-13}$ and 
$1.5\times10^{-12}$\,\Msun/yr. Factoring the broad range of ages derived from 
various stellar evolution models of 200 -- 250 Myr \citep{2017ApJ...840...70B,2005ApJ...630L..69L}, the total mass loss for 
Sirius A is significantly smaller than the uncertainties in mass determinations 
from astrometry. For these reasons, we have decided to use a fixed mass of 
$2.06$\,\Msun~for Sirius A, in our model grid. 

One of the goals of our study is to better constrain the initial mass of Sirius 
B. \cite{2018ApJ...866...21C} derived the IFMR based on self-consistent analysis 
of Sirius B and 79 white dwarfs from 13 clusters. Based on that study, and also 
considering previous IFMR studies, there is a rather large uncertainty in 
constraining the initial value masses. The current mass of Sirius B is 
$1.018\pm0.011$\,\Msun, which could have been from a  progenitor with initial 
mass from 4 to 6.8\,\Msun. We have calculated our model grid with initial masses 
for Sirius B in that range, with 0.1\,\Msun~steps.  

\subsection{Photospheric parameters} \label{spec}

With V magnitudes of -1.47 \citep{1953ApJ...117..313J} and 8.44 
\citep{1998ApJ...497..935H} for Sirius A and B, respectively, therefore very distinct luminosities, this binary system
imposed a puzzle for having comparable spectral energy distribution. 
The effective temperature of Sirius B is \teff\, = $25193 \pm 37$\,K \citep{2005MNRAS.362.1134B} and  \teff\, $=26083\pm378$\,K 
\citep{2017ApJ...848...11B} from HST and ground-based optical spectroscopy respectively. 
We have used \teff~ between 25000 and 26000 for Sirius B as a guide to establish the stopping condition of our calculated models. 
\cite{2017ApJ...840...70B} quotes the luminosity to be $0.02448 \pm 0.00033$\,L$_\odot$, but using the Gaia magnitude $G=8.52$ and parallax $\Pi=374.49\pm0.23$\,\arcsec, we calculated log$(L/\mathrm{L}_\odot) = -1.54$, by taking into account bolometric corrections. 

\cite{2011PASA...28...58D} used interferometry to determine the fundamental quantities of Sirius A. Through observations made at 694.1\,nm, they determined the radius to be $1.713 \pm  0.009$\,R$_\odot$ by combining their measurements of the angular size with the Hipparcos parallax. They also determined the bolometric flux to be $F=(5.32 \pm 0.14)\times 10^8$\,W/m$^2$, the effective temperature and luminosity values yielded are \teff\, = $9845 \pm 64$\,K and $L=24.7 \pm 0.7$\,L$_\odot$ respectively. 
We have used the luminosity and \teff~for Sirius A to constrain the initial metallicity of the system. 

\cite{2011A&A...532L..13P} measured a constant, weak magnetic field of $0.2\pm0.1$\,G for Sirius A. Because there are no measurements of magnetic field for Sirius B, we have used this as the initial value for Sirius A and B. There is no evidence that both stars would have been born with different magnetic fields. 

\subsection{Metallicity} \label{fe}

\cite{1925PhDT.........1P} made the first observational 
study of chemical abundances in stars, measuring the 
fraction of heavy elements in the universe. In principle, 
the abundance of any element other than helium can be 
compared to that of hydrogen to determine metallicity. 
The elemental abundance of a star can be used to 
empirically determine the age of the system, as younger 
stars will generally be more metal rich 
\citep{2010ARA&A..48..581S}. Metallicity can also provide 
insight as to how a star is evolving, as the isochrones 
of a population of stars and its evolutionary 
track can shift due to changes in its metal content 
\citep{2012MNRAS.427..127B}.

Modeling stellar evolution is dependent on the chemical abundance of the progenitor stars. For binary systems, we can assume that both stars have similar initial metallicities. 
Specifically for the Sirius system, since we cannot determine the original metallicity for Sirius B, we constrain its value based on the companion's on the main sequence, Sirius A \citep{2021ApJS..253...58Q}. 

There have been multiple determinations of Sirius A's chemical abundances [e.g.\citep{2016ApJ...826..158C,2011A&A...528A.132L}]. Regardless the scatter in the independent determinations and the analysis of different elements, all results point towards high metallicity values. However, previous attempts to model the stellar evolution of Sirius A cannot find high metallicity solutions compatible with the observed luminosity and the best fits for \teff. \cite{2017ApJ...840...70B} computed evolutionary models using TYCHO and YREC codes and determined the best model to have a sub-solar metallicity of $Z=0.85$\,Z$_\odot$. It is important to note that the metallicity in the models is referred to as the initial value. 

As an input in our models, we tested a broad range of metallicities, as the current values for Sirius A could have been somewhat contaminated by the planetary nebula of Sirius B. 

\subsection{Past binary interaction} \label{interact}

There are many reasons that support a lack of an interacting past, such as the rotation rate of Sirius A compared to other stars in its spectral class, their current separation, amongst others \citep{2017ApJ...840...70B}. \cite{2016ApJ...826..158C} found evidence of copper and other heavy elements present in the spectra of Sirius A, indicating that there could have been accretion in the past due to the presence of s process elements. \cite{2017ApJ...840...70B} created a grid using TYCHO to determine if accretion influenced the evolution of Sirius A. They compiled models starting with a subsolar metallicity of $0.9$\,Z$_\odot$ and an initial mass of 1.96\,\Msun~and evolved the system until the AGB phase, right before pulsation, for 100\,MYr \citep{2017ApJ...840...70B}. This, however, resulted in a radius that is too large for the current luminosity of Sirius A and an overall worse fit of the observed stellar parameters.

Based on the previous studies and observations, we have constructed our model grid assuming that the stars did not develop a common envelope nor that mass transfer happened in any stages of the past evolution of the system. 

\section{Computational Methods} \label{mesa}

MESA, or Modules for Experiments in Stellar Astrophysics, is an open-source code that allows users to model stellar structures and evolution \citep{2011ApJS..192....3P, 2013ApJS..208....4P, 2015ApJS..220...15P, 2018ApJS..234...34P, 2019ApJS..243...10P, 2023ApJS..265...15J}. MESA uses numerical methods to simultaneously solve the coupled, non-linear differential equations for stellar evolution in a modular structure with a collection of
Fortran libraries. 
MESA is used over a broad range of subjects, as it allows users the freedom in choosing prescriptions for various physical parameters (e.g. mass, wind, opacity, etc.). Besides single stars, MESA can be used to compute the evolution of binary systems. 

We have used most recent release of MESA (version 23.05.1) to compute our model grid. 
We have used the binary capabilities of MESA (\texttt{evolve\_both\_stars} binary module) to model the Sirius system from the pre-main sequence up until the current observed evolutionary stage. MESA simultaneously evolves an interacting binary system, accounting for mass exchange, accretion, and changes to the orbital mechanics \citep{2015ApJS..220...15P}.

Some of the assumptions this module makes is that the rotational axis is completely perpendicular to the orbital plane, which is a good enough approximation for most physical situations \citep{2015ApJS..220...15P}. It can also account for the tidal interaction forces resulting from accretion, though this feature was not used in this analysis. The binary module is constructed with the Kolb and Ritter mass loss schemes, both of which are defined in \cite{2015ApJS..220...15P}.

The second module that we used was \texttt{make\_co\_wd} to create a white dwarf with an even distribution of carbon and oxygen in the core, with a hydrogen and helium photosphere from a specified main sequence star \citep{2023ApJS..265...15J}. It is relevant to mention that this module omits the thermally pulsating phase \citep{2023ApJS..265...15J}. 

\section{MESA Model Grid}\label{grid}

We computed stellar evolutionary models using MESA, 
evolving the binary system from the zero age main 
sequence until the current location of Sirius B in the white dwarf cooling sequence. As discussed in the previous sections, we 
have fixed the mass of Sirius A to be 2.063\,\Msun. We have also fixed the value for magnetic field to be $B=0.2$\,G. We have used the values for observed luminosities and best determinations of \teff~for Sirius A and B as guide for stopping conditions. This tells us that the metallicity values that make the model fit the spectroscopically observed parameters yield an accurate evolution. The MESA inlists and example models can be found via this \dataset[hyperlink]{https://zenodo.org/records/13914550}.

\subsection{Testing Metallicity}\label{test_fe}

As discussed in section~\ref{fe}, we assume that the original metallicity of Sirius B is the same as for Sirius A. We started our calculations by using previous determinations of metallicity for Sirius A. We have computed evolutionary models in attempts to fit the observed metallicity of Sirius A. 

\begin{figure}[h!]
\includegraphics[scale=0.5]{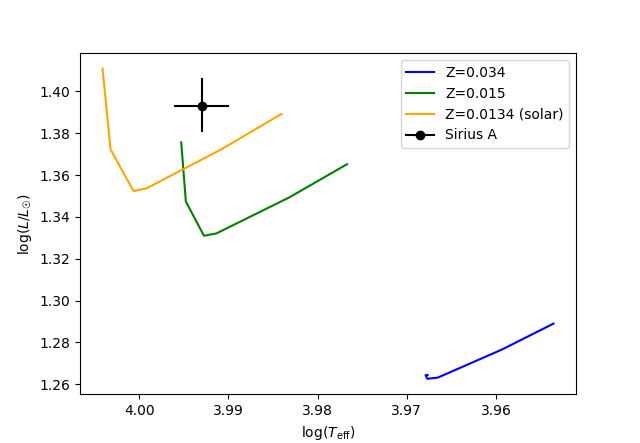}
\caption{HR diagram for the evolutionary tracks of Sirius A for a high metallicity of $[\mathrm{Fe}/\mathrm{H}]=0.4$ (or $Z=0.034$) determined by \cite{2018MNRAS.477.3343S} (blue line), solar metallicity (orange line), and another model with super-solar of $Z=0.015$ estimated from \cite{2016ApJ...826..158C} (green line), in comparison to the observed luminosity and \teff. None of the high metallicity models is in agreement with the current position of Sirius A in the HR diagram.
\label{metals_AA}}
\end{figure}

In Figure~\ref{metals_AA}, we plot our calculated evolutionary tracks of Sirius A for high \citep{2018MNRAS.477.3343S}, super-solar \citep{2016ApJ...826..158C}, and solar metallicities
compared to the observed luminosity and \teff. None of our models are in agreement with the current observations of Sirius A. This discrepancy could be explained by the fact that the input values in the models are the initial metallicity and/or by possible contamination from the Sirius B planetary nebula. 

\begin{figure}[h!]
\includegraphics[scale=0.45]{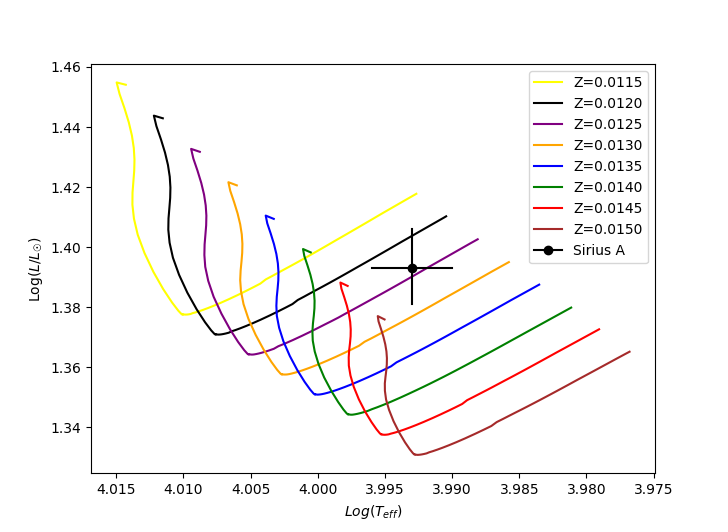}
\caption{HR diagram for the evolutionary tracks calculated with MESA for Sirius A. 
We have used different initial metallicities from $Z=0.0115$ (yellow line, top left) to 0.0150 (dark red, bottom right), in steps of 0.0005. Compared to the observed luminosity and determinations of \teff, our best fit is a subsolar metallicity of $Z=0.0124\pm^{0.00073}_{0.00077}$. 
\label{metals_A}}
\end{figure}

We then proceeded to compute models with various metallicities from $Z=0.0115$ to 0.0150, in steps of 0.0005. In Figure~\ref{metals_A}, we plot evolutionary models for Sirius A, compared to the observed luminosity and measured \teff. The best fit for initial metallicity of Sirius A is $Z=0.0124\pm^{0.00073}_{0.00077}$, a subsolar value ($Z=0.93\,Z_\odot$), in agreement with previous calculations using TYCHO and YREC \citep{2017ApJ...840...70B}. We have used this value as the initial metallicity for both stars in the Sirius system.

\subsection{Mass of Sirius B}

One of our main goals in computing evolutionary models is to better constrain the mass of the progenitor of Sirius B. We have varied the 
initial mass from 4.0\,\Msun~to 6.8\,\Msun, in 
steps of 0.1\,\Msun, using the luminosity, spectroscopic \teff, and astrometric mass of Sirius B 
as a guide for stopping condition, and implementing the best fit for the metallicity as an initial condition, from the analysis on Sirius A. 

\subsection{Other parameters into the models}

We computed our models with a mesh delta coefficient of 3.75. This coefficient is related to how incrementally the code splits up the star. At each of the boundaries created by this mesh, the equations of stellar evolution are solved and made to be continuous across the boundary. Mass overflow was monitored by the Roche Lobe overflow limit, which determines if accretion would occur, making the binary stars interact in a significant manner. For the Sirius system, the orbital solution from astrometry places the stars too far apart for mass transfer. 

We modeled the progenitor of Sirius B taking into account wind mass loss effects. We selected the MESA's wind mass loss function that computes stellar angular momentum while taking into account the removed layers' momentum \citep{2011ApJS..192....3P}. 

\section{Results}\label{results}

We have computed evolutionary models for both stars simultaneously, assuming that the initial mass for Sirius A as $M=2.063$\,\Msun, non-interacting stars, and neglecting mass loss for Sirius A. In Table~\ref{initial_par}, we summarize the initial parameters of our model grid. 

\begin{table}[h]
\centering
\begin{tabular}{|c|c|c|c|c|c|}
\hline
& $M$\,(\Msun) & $Z$ & $\log L/\mathrm{L}_\odot$ & $\log$ \teff & $B$\,(G) \\ \hline\hline
 A & 2.063 & 0.0124 & 1.37 -- 1.39 & 3.993 -- 4.001 & 0.2 \\ \hline
 B & 4 -- \textbf{6.8} & 0.0124 & -1.62 -- -1.54 & 4.398 -- 4.433 & 0.2 \\ \hline
\end{tabular}
\caption{Fundamental stellar parameters used to calculate models with MESA \texttt{evolve\_both\_stars}. We assumed that there was no mass transfer between the two stars and that the mass of Sirius A stays constant throughout its time on the main sequence, because of the negligible mass loss. This temperature range and luminosity were used to place stopping conditions in the evolutionary models.}
\label{initial_par}
\end{table}

In Table ~\ref{bestfittable}, we list the values of our model grid calculations for Sirius B: initial and final masses, age, and \teff. 
The metallicities used for the analysis of Sirius B were based on the values for Sirius A that best fit the observed luminosity and spectroscopic \teff, in the  range $Z=0.0124\pm^{0.00073}_{0.00077}$. %The luminosity was used as guide for the stopping condition in the white dwarf cooling sequence. 

\begin{table}[h!]
    \centering
    \begin{tabular}{|c|c|c|c|}
    \hline
$M_{\mathrm{initial}}$\,(\Msun) & $M_{\mathrm{final}}$\,(\Msun) & Age\,($10^8$\,years) & $\log$(\teff) \\ [0.5ex] 
      \hline\hline
        4.0&0.839& 3.006  & 4.369 \\
               \hline
        4.1 &0.855& 2.905& 4.367 \\
         \hline
        4.2 &0.861& 2.803 & 4.372 \\
         \hline
        4.3 &0.869& 2.714 & 4.375  \\
         \hline
        4.4 &0.873 & 2.663 & 4.377  \\
         \hline
        4.5&0.880 & 2.550  & 4.379  \\
         \hline
        4.6&0.889 & 2.483  & 4.381  \\
         \hline
        4.7&0.900& 2.430 & 4.376 \\
         \hline 
        4.8&0.888& 2.372& 4.379\\
         \hline
        4.9&0.894& 2.322  & 4.381  \\
         \hline
        5.0&0.901& 2.275  &4.390\\
         \hline
        5.1&0.907& 2.227 &4.385 \\
         \hline
        5.2&0.914 & 2.198 & 4.396  \\
         \hline
        5.3&0.949& 2.169 & 4.401 \\
         \hline
        5.4&0.930& 2.139  & 4.401  \\
         \hline
        5.5&0.970& 2.121  & 4.406  \\
  \hline5.6&0.945& 2.071  & 4.409  \\
  \hline5.7&0.955& 2.050  & 4.415  \\
         \hline
        5.8&0.964& 2.016  & 4.411  \\
         \hline 
        5.9&1.039& 1.834  & 4.443  \\
         \hline
        6.0&1.015& 2.036  & 4.414  \\
         \hline
        
                 6.1& 1.047&  2.114  &  4.420  \\
         \hline
                  6.2& 1.038&  2.008  &  4.424  \\
         \hline
                  6.3& 1.092&  1.882  &  4.444  \\
         \hline
                  6.4& 1.115&  2.138  &  4.442  \\
         \hline
                  6.5& 1.142&  2.187  &  4.459  \\
         \hline
                  6.6& 1.147&  1.939  &  4.467  \\
         \hline
                  6.7& 1.155&  1.989  &  4.459  \\
         \hline
                  6.8& 1.168&  2.146  &  4.468  \\
         \hline
         
    \end{tabular}
    \caption{Table or models calculated for Sirius B at the metallicity of $Z=0.0124$, for different initial and final masses, ages, and \teff.}
    \label{bestfittable}
\end{table}

\begin{figure}
    \centering
    \includegraphics[scale=.5]{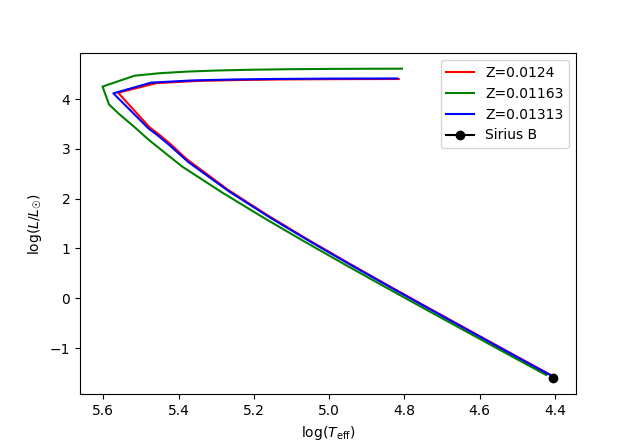}
    \caption{HR Diagram for Sirius B for the best fit metallicity, including the upper and lower bounds, for a progenitor mass of $6.0\,\mathrm{M}_\odot$. The white dwarf cooling track agrees with the current placement of Sirius B on the luminosity temperature plane.}
    \label{siriusbhrgrid}
\end{figure}

In Figure~\ref{siriusbhrgrid}, we plot the HR diagrams showing models for the white dwarf cooling sequence, for a fixed progenitor mass of 6.0\,\Msun, for the upper and lower limits in metallicity, determined from Sirius A. For white dwarf cooling, changes in metallicity produce minute changes in the evolutionary tracks. 

%\subsection{Determining the Best Fit}

Once the models were calculated, we searched in our grid for the solutions closest to the mass $(M)$ from astrometry \citep{2017ApJ...840...70B}, effective temperature (\teff) from spectroscopy \citep{2005MNRAS.362.1134B, 2017ApJ...848...11B}, and luminosity $(L)$ from astrometry and parallax \citep{2017ApJ...840...70B}. We minimized the squared of the differences of the normalized physical quantities $(\xi_i)$ listed above ($M$, \teff, $L$) from the models and the observations as:

\begin{equation}
\chi = \sqrt{\sum_i\left(\frac{\xi_{i,\mathrm{observed}}-\xi_{i,\mathrm{model}}}{\xi_{i,\mathrm{observed}}}\right)^2}
\end{equation}

\begin{figure}
    \centering
    \includegraphics[scale=0.5]{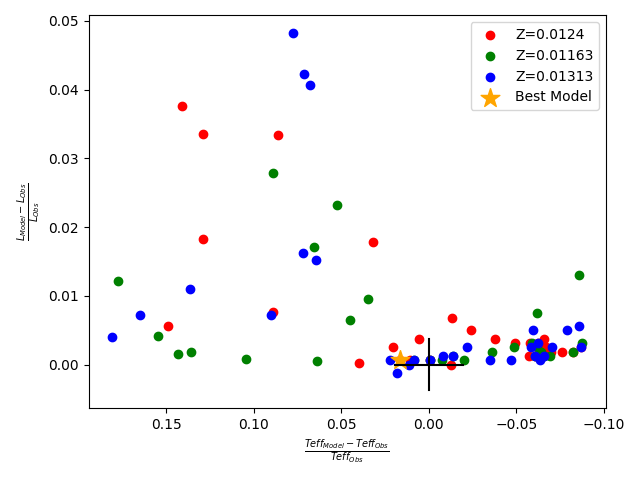}
    \caption{Comparison between models and the observations of Sirius B in $L$--\teff~plane. The best model (yellow star) was found by the minimization of the normalized luminosity, spectroscopic \teff, and astrometric mass (not displayed in this plot). The models are for the calculated metallicity (red dots) and upper and lower limits (blue and green dots, respectively. The spread of the models becomes large for progenitor masses over $6.2\,\mathrm{M}_\odot$.} 
    \label{lumtemp}
\end{figure}

\begin{figure}
    \centering
    \includegraphics[scale=0.5]{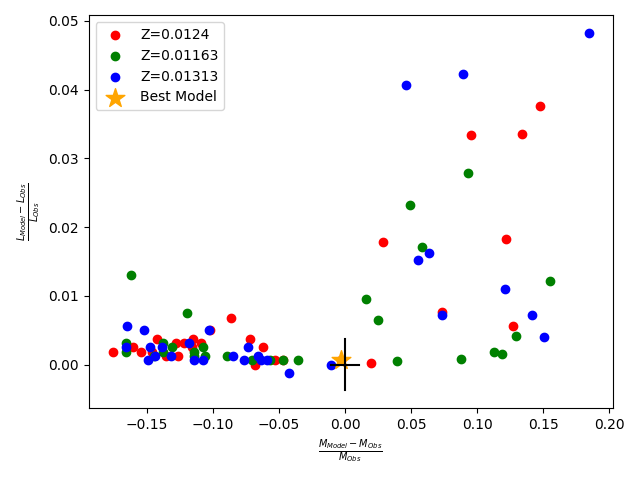}
    \caption{Comparison between models and the observations of Sirius B in $L$--$M$~plane. Just as before, The best model (yellow star) was found by the minimization of the normalized luminosity, spectroscopic \teff, and astrometric mass. This plot highlights that smaller initial masses of Sirius B do not fit the spectroscopic parameters very well. With this, we can put a lower limit on acceptable progenitor masses.}
    \label{lummass}
\end{figure}

\begin{figure}
    \centering
    \includegraphics[scale=0.5]{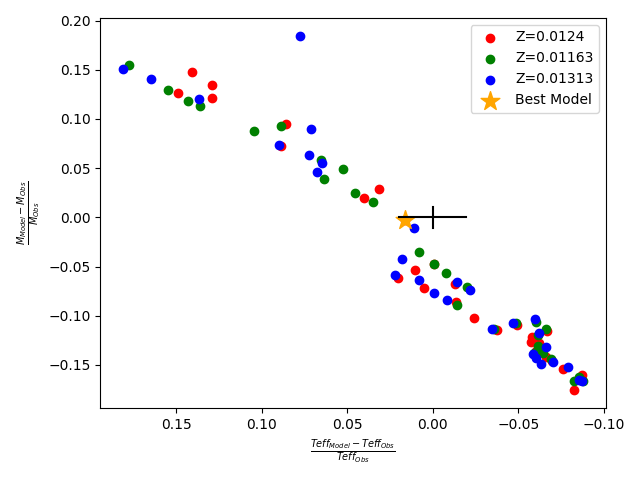}
    \caption{Comparison between models and the observations of Sirius B in $M$--\teff~plane. Despite there being many models that fall into the range of acceptable temperature, the mass is so well constrained that it limits the possible solutions.}  
    \label{masstemp}
\end{figure}

In Fig.~\ref{lumtemp}, we plot our models and the minimized observed quantities of \teff~and luminosity. We also use mass as a third constraint to put limits on viable models. Fig.~\ref{lummass} and Fig.~\ref{masstemp} display the other views of this three-dimensional space. 
Our best solution for initial mass of Sirius B and the age of the system are $6.0\,\mathrm{M}_\odot$ and 203.6\,Myr, respectively.
By using the uncertainties in luminosity, spectroscopic \teff~and astrometric mass, we computed that the uncertainties in the age of the system is $\pm 45$\,Myr, and in progenitor mass for Sirius B is $\pm 0.6$\,\Msun. 
Additionally, our models indicated an \teff=$25882 \pm 2332$\,K for Sirius B.  

%the solutions were constrained by placing them in the mass and temperature plane. The results are plotted in Figure~\ref{familyofsolutions}, displaying the trend of increasing final white dwarf mass yielding hotter temperatures. By determining the minimum distance for a solution to the well-established white dwarf mass \citep{2017ApJ...840...70B} of $1.018 \pm 0.011$ and an effective temperature of $25970 \pm 380$ determined by \cite{2017MNRAS.465.2849T}. The temperature and mass values from the models were normalized to the spectroscopic and astrometric observational values, respectively. We then minimized the differences between the observations and the models. Our best solution is for a progenitor mass of $6.0$\,\Msun for Sirius B with a metallicity of $Z=0.0124$, yielding a total age of the binary system from ZAMS to the current evolutionary phase to be 203.6\,Myr. This age was ammended from the originally obtained value of 185 Myr, indicating that the actual luminosity required to fit Sirius B on the HR mass/temperature plane is on the lower end of the spectroscopic limit.

\section{Comparison to Previous Determinations}

%\subsection{White Dwarf Cooling Tracks}

We now turn to review how our determinations compare with results from previous studies. 

In addition to the orbital solution and mass determinations for the Sirius system, \cite{2017ApJ...840...70B} modelled the evolution of the stars, using the ``Montreal'' photometric tables for white dwarf cooling sequences \citep{2006AJ....132.1221H}, combined with PARSEC \citep{2012MNRAS.427..127B} and YREC evolutionary models \citep{2008Ap&SS.316...31D}. They have determined the total age of Sirius B is $228\pm10$\,Myr, for a cooling age of $\sim126$\,Myr. For Sirius A, using a slightly subsolar metallicity of about $0.85\,\mathrm{Z}_\odot$, they found that the best models yield ages between 237 and 247\,Myr, with uncertainties of $\pm15$\,Myr, consistent with that of the white dwarf companion. In a previous study, \cite{2005ApJ...630L..69L} using similar model grids determined the age of Sirius system to be 225--250\,Myr. Our best solution for age is $203.6\pm45$\,Myr, which within $2\sigma$ from the other determinations, or 10\% difference. 

This discrepancy in age determination can be explained from different prescription in the physics and choices in physical quantities
input in the models. Obviously, mass is the most important parameter, which dictates stellar evolution. \cite{2017ApJ...840...70B} 
estimated the progenitor mass for Sirius B to be $5.6\pm0.6$\,\Msun\, 
from PARSEC, and $5.0\pm0.2$\,\Msun\, from YREC models. Our models fit 
best a progenitor with $6.0\pm0.6$\,\Msun, which makes the Sirius 
system younger than previously determined. From our 
Table~\ref{bestfittable}, the 225\,Myr model would be for the a 
star with an initial mass of $5.05$\,\Msun, consistent with the previous determinations. However, the final mass is $0.904$\,\Msun, which is smaller than the astrometric mass and off by $10\sigma$. Since the final mass of Sirius B is one of the most reliable measurements for this system, we can state that using different evolutionary codes will \textbf{yield} external uncertainties of about $20$\,Myr in age, $1$\,\Msun\, in initial mass, and $0.1$\,\Msun\, in final mass.

Metallicity also affects the evolutionary timescales in the models, as stars with higher percentage of metals have a shorter lifetime. \cite{2017ApJ...840...70B} used in their evolutionary model calculations a metallicity of $Z=0.85$\,Z$_\odot$, while our best fit is for $Z=0.92^{0.06}_{0.05}$\,Z$_\odot$, both subsolar values. Because our determination is for a larger metallicity, our age determination, as expected, is a younger system.

Finally, one of our main goals with this study is to contribute in constraining of the IFMR. \cite{2018ApJ...866...21C} determined the IFMR by creating isochrones for a dataset of clusters and binary stars using PARSEC and MIST non-rotating isochrones.

However due to the large initial mass of our solution, the IFMR is smaller in comparison to the PARSEC and MIST determinations with a value of $0.169$ $M_\mathrm{final}/M_\mathrm{initial}$ versus $0.177$ $M_\mathrm{final}/M_\mathrm{initial}$ for a $6.0$\,\Msun\ progenitor \citep{2018ApJ...866...21C}.

Our best fit for the initial and final mass for Sirius B contributes into the spread in determinations, due to the higher metallicity. 

%The spread in the IFMR are a result of fitting continuous three-piece relations as well as a systematic difference between observed white dwarf masses and the theoretical counterparts for intermediate and high mass stars \citep{2018ApJ...866...21C}. Additionally, the uncertainty stems from effects from metallicity, stellar environment, and also how age of the system is defined differently between single stars, binaries, and clusters \citep{2016ApJ...818...84C}. We calculate our own IFMR for Sirius B and compare the results with the aforementioned isochrones in figure ~\ref{ifmr} for our best fit metallicity of $Z=0.0124$ including the spread in the final mass due to chemical composition. 

%\begin{figure}[h!]
    %\centering
 %   \includegraphics[scale=0.5]{Plots/fixed plots/IFMR0124.png}
  %  \caption{Plotting the initial to final mass ratio of the Sirius system grid calculated by MESA for $Z=0.0124$. Comparing the results to both the Mist and PARSEC codes. The uncertainties become exceedingly larger the bigger the initial mass of the progenitor becomes. Even with initial conditions remaining the same, there is a spread in the white dwarf mass due to how MESA interpolates for a given metallicity.}
   % \label{ifmr}
%\end{figure}

\section{Conclusion and Future Works}

In this paper, we report our best fit solutions to model the Sirius system, by using MESA models, constraining the parameter space with the observational physical quantities. We discussed our levels of confidence and the reasoning of why some quantities place stronger constraints than others. Our best model for metallicity of Sirius A is $Z=0.0124$, and using a fixed mass of $2.063$\,\Msun\, from astrometry and Sirius B's luminosity, \teff, and final mass, yields a progenitor mass of $6.0 \pm 0.6$\,\Msun. 
This is in agreement with \cite{2017ApJ...840...70B}, the main differences being our larger progenitor mass for Sirius B and the notably lower age of the system. Using the PARSEC and Tycho IFMR's, \cite{2017ApJ...840...70B} obtained initial masses of $5.6 \pm 0.6$\,\Msun and $5.0 \pm 0.2$\,\Msun. Our results fall into the range of the PARSEC model, but the values determined with the YREC code are significantly smaller, in part due to the lower metallicity chosen by \cite{2017ApJ...840...70B} and possibly also the treatment of the thermal pulsations during the ABG phase. 

We have used a non-interactive model for the Sirius system, based on the current orbital solution from previous studies. The fact that we could not fit the current observed high metallicity for Sirius A remains a puzzle, but we argue that the state-of-the-art models only take into account substantial interactions between binary stars and cannot account for minor pollution from the planetary nebula.

There are limitations in our calculations due to assumptions and choices in parameters for the calculations of the models. We must point out that we have not changed the time step nor the mesh point between each model. However, we have presented a systematic approach to modeling evolution of binary systems, constraining the most reliable physical parameters, while varying the least precise ones within the uncertainties. We have also discussed the external uncertainties between our determinations and the ones from previous studies. We are encouraged by our results to expand the parameter space in our calculations since our results are consistent with the ones derived using other model grids. We will then attempt to model other binary systems, starting with ``Sirius-like'' systems and expanding into interacting systems.

%\begin{acknowledgments}

%\end{acknowledgments}

%% For this sample we use BibTeX plus aasjournals.bst to generate the
%% the bibliography. The sample631.bib file was populated from ADS. To
%% get the citations to show in the compiled file do the following:
%%
%% pdflatex sample631.tex
%% bibtext sample631
%% pdflatex sample631.tex
%% pdflatex sample631.tex

\bibliography{sirius_system}
\bibliographystyle{aasjournal}

\end{document}